\documentclass[aps,pra,reprint,groupedaddress,showkeys]{revtex4-2}

\usepackage{graphicx}
\usepackage[colorlinks=true,citecolor=blue,urlcolor=blue]{hyperref}
\begin{document}

\title{Effect of niobium doping on the local structure \\
and oxidation state of Co atoms in SrTiO$_3$}

\author{A. I. Lebedev}
\email[]{swan@scon155.phys.msu.ru}
\author{I. A. Sluchinskaya}
\affiliation{Physics Department, Lomonosov Moscow State University, 119991 Moscow, Russia}


\begin{abstract}
XAFS spectroscopy combined with first-principles calculations are used to study SrTiO$_3$ samples
simultaneously doped with cobalt and niobium. It is shown that additional doping of the samples
with Nb donors results in a decrease in the oxidation state of cobalt atoms to +2, which is
accompanied by an increase in the Co--O distance to the nearest oxygen atoms, in agreement with
first-principles calculations. The impurity Co atoms are shown to occupy Ti sites and are on-center.
The suppression of the oxygen-vacancy formation upon doping SrTiO$_3$(Co) with niobium was
revealed. First-principles calculations indicate that the Coulomb interaction of ions with local
electric fields generated by non-isoelectronic Nb impurities plays a decisive role in the local
distortions of atomic positions observed around the cobalt atom in the unit cell. The possibility
of controlling the magnetic moment of Co atoms by intentional change in the lattice parameter
is demonstrated.
\end{abstract}


\widetext{\texttt{Authors' translation of the paper published in Fizika Tverdogo Tela {\bf 68},
882 (June 2026).}}

\maketitle

\section{Introduction}

Owing to their unique properties, strontium titanate and its solid solutions are of significant
interest for both fundamental research and practical applications, making them the subject of
intensive study~\cite{SrTiO3-book}. Doping SrTiO$_3$ with $3d$ elements enables the modification
of its dielectric, electrical, optical, and magnetic properties, thereby broadening its
potential applications in microelectronics.

When SrTiO$_3$ is doped with cobalt, which substitutes for Ti at the $B$~site of the perovskite
structure, the emergence of room-temperature ferromagnetism was discovered. SrTiO$_3$(Co)
thin films, grown by pulsed laser deposition and co-doped with the La donor impurity to
achieve metallic conductivity, exhibit ferromagnetism at 300~K when the cobalt concentration
is between 1.5\% and 2\% and the films are annealed in a reducing
atmosphere~\cite{ApplPhysLett.83.2199,PhysRevLett.96.027207}. Later, ferromagnetism at 300~K
was reported in semiconducting cobalt-doped SrTiO$_3$ thin films grown by pulsed laser
deposition or molecular beam epitaxy at impurity concentrations exceeding
14\%~\cite{NewJPhys.12.043044,PhysRevB.87.144422}. Conversely, SrTiO$_3$ films doped with
2\% Co and co-doped with 0.5--1\% Nb showed no evidence of ferromagnetism down to
5~K ~\cite{ApplPhysLett.89.012501}. Theoretical studies~\cite{PhysRevB.87.144422,
ApplPhysLett.100.252904,PhysRevB.90.125130} suggest that oxygen vacancies, in addition to
magnetic Co impurities, play a significant role in the emergence of ferromagnetism in
SrTiO$_3$; these vacancies are inevitably formed to compensate for the difference in charge
states between Co and Ti ions.

The interest to study Co-doped SrTiO$_3$ is also driven by its extensive use as an ionic
conductor. Although undoped strontium titanate is not an ionic conductor, its doping with
specific impurities---namely cobalt---generates oxygen vacancies that electrically compensate
for the valence difference between the Ti and the dopant, thereby transforming the material
into an oxygen-ion conductor. With a cobalt concentration of 25--35\%, this material shows
promise for applications in oxygen-conducting membranes, which are systems designed to extract
oxygen from air or from mixtures of oxygen with other gases~\cite{JEurCeramSoc.38.5058}.

The SrTiO$_3$(Co) ionic conductor doped with cobalt only is also a hole conductor, exhibiting 
a mixed hole--ionic conductivity. Compared to iron---a common material in fuel cell
development---cobalt offers a higher catalytic activity, thus enabling operation at lower
temperatures. When co-doped with donors like niobium, SrTiO$_3$(Co) is suitable for fuel
cell anodes~\cite{IntJHydrogenEnergy.34.6407} and cathodes~\cite{ECSTrans.91.1291}, provided
the cobalt and donor concentrations are carefully optimized to maintain chemical stability
in both oxidizing and reducing environments. Besides, nanocrystalline SrTiO$_3$(Co) has proven
to be highly effective for hydrogen production through photoelectrochemical water
splitting~\cite{EnvironProgSustainEnergy.42.e14186}.

And finally, magnetic $3d$ elements can occupy the $A$~site of the perovskite structure and
become off-center, as was previously shown for Mn~\cite{ApplPhysLett.86.172902,JETPLett.89.457,
BullRASPhys.74.1235}, Fe~\cite{PhysSolidState.64.345,Ferroelectrics.604.116}, and
Co~\cite{PhysSolidState.61.390,JAlloysComp.820.153243} impurities. In this case, SrTiO$_3$ can
acquire the properties of a magnetoelectric glass, in which disordered polar and magnetic states
emerging at low temperatures coexist and give rise to magnetoelectric interaction. Such a state
was previously observed in Mn-doped SrTiO$_3$~\cite{PhysRevLett.101.165704}. This opens up
the possibility of controlling the properties of these materials with a magnetic field.

The incorporation of Nb into strontium titanate is of interest in itself because, upon heavy
doping with niobium ($\sim$20\%), the material acquires a relatively high thermoelectric
figure of merit ($ZT = 0.37$)~\cite{ApplPhysLett.87.092108}.

The properties of cobalt-doped SrTiO$_3$ have been described in detail in our earlier
publications~\cite{PhysSolidState.61.390,JAlloysComp.820.153243}. In those papers, XAFS
spectroscopy was used to investigate the local structure and oxidation state of the
Co impurity in SrTiO$_3$. It was shown that cobalt can occupy both sites in the
perovskite structure, with oxidation states of +2 at the $A$~site and +3 at the $B$~site.
It was also revealed that Co is an off-center atom at the $A$~site, displaced by approximately
1.0~{\AA} from the lattice site. First-principles calculations demonstrated that an isolated
Co$^{3+}$ ion at the $B$~site is diamagnetic, whereas the Co$^{2+}$ ion at the $A$~site is
in a high-spin state ($S = 3/2$). By varying the synthesis conditions, the ratio of cobalt
concentrations at the $A$ and $B$~sites can be deliberately controlled. In particular,
synthesis conditions were found under which cobalt in SrTiO$_3$(Co) samples predominantly
substitutes at the $A$~site.

To understand the phenomena observed in strontium titanate doped with cobalt and compensated
by niobium, it is necessary to know the microscopic structure of the resulting defects.
In particular, it is important to understand the role of Nb impurities in the evolution of
structural defects in SrTiO$_3$(Co) and to find ways to selectively control the position,
oxidation state, and magnetic state of Co atoms. This information can be obtained by studying
samples using XAFS spectroscopy in conjunction with first-principles modeling of the defect
structure, thereby enabling more reliable identification of the defects formed.

\section{Samples and used techniques}

SrTiO$_3$ samples with a strontium-enriched nonstoichiometric composition, doped with 3\% cobalt,
as well as samples with the same Co concentration and co-doped with 6\% Nb, were prepared by
solid-state synthesis. The starting materials were SrCO$_3$ and nanocrystalline TiO$_2$
(obtained by the hydrolysis of tetrapropylorthotitanate and dried at 500$^\circ$C).
Co(NO$_3$)$_2$$\cdot$6H$_2$O and a pre-synthesized CoNb$_2$O$_6$ columbite phase, prepared from
Co$_3$O$_4$ and Nb$_2$O$_5$ at 1100$^\circ$C, served as dopants. The components were weighed
in the required proportions, ground in acetone using an agate mortar, and annealed in alumina
crucibles at 1100$^\circ$C for 4~hours in air. The resulting powders were ground again and
re-annealed under the same conditions for another 4~hours. Some samples were additionally
annealed at 1500$^\circ$C for 2~hours in air. All the obtained samples were dark brown in color.
Upon the introduction of niobium, the dark brown color of the SrTiO$_3$(Co) powder became lighter,
taking on an orange hue.

X-ray absorption spectra in the extended fine structure (EXAFS) and near-edge structure (XANES)
regions were recorded in fluorescence mode at the Co $K$ absorption edge (7.709~keV) at 300~K at
the KMC-2 beamline of the BESSY synchrotron radiation source (Germany). After annealing, the
samples were ground again and applied to the surface of adhesive tape. The sample thickness
required for optimal recording of spectra was achieved by repeatedly folding the tape (typically
8~layers were required). The radiation was monochromatized using a (111)-oriented
Si$_{1-x}$Ge$_x$ double-crystal monochromator. The intensity of the radiation incident on the
sample was measured using an ionization chamber, while the fluorescence intensity was measured
using a R{\"O}NTEC silicon drift detector operating in energy-dispersive mode. XAFS spectra were
processed using the IFEFFIT software package~\cite{JAmChemSoc.113.5135}. Measurements on
CoNb$_2$O$_6$ and LaCoO$_3$ reference compounds, in which Co atoms share a similar local environment
(i.e., the centers of octahedra), were performed in the same way. Several spectra were recorded
for each sample and processed independently; the resulting $\chi(k)$ curves were then averaged.
Because of the low impurity concentration in the samples, no correction for self-absorption of
the fluorescence radiation was applied. Details of the data processing are provided in~\cite{PhysRevB.55.14770}.

Calculations of the geometry, magnetic structure, and electronic structure of cobalt-containing
impurity centers were performed from first principles using the ABINIT code~\cite{ComputPhysCommun.205.106}
within the LDA+$U$ approximation, using 80-atom cubic supercells in which one Ti$^{4+}$ ion at the
$B$~site of the perovskite structure was replaced by a cobalt ion. To modify the oxidation state of
cobalt, acceptor (Rb$_{\rm Sr}$) or donor (Nb$_{\rm Ti}$, F$_{\rm O}$) impurities located as far
as possible from the Co atom were introduced into the supercell. Donor oxygen vacancies located
either in the immediate vicinity of the Co atom or at a distance from it were also considered. PAW
pseudopotentials~\cite{ComputMaterSci.81.446} were used to describe atoms with partially filled
$d$~shells. The parameters describing the intra-$d$-shell Coulomb and exchange interactions were
set to $U = 5$~eV and $J = 0.9$~eV.

\section{Results}

\subsection{X-ray measurements}

X-ray phase analysis of SrTi$_{0.97}$Co$_{0.03}$O$_3$ samples doped only with cobalt and annealed
at 1100 and 1500$^\circ$C showed that all samples were single-phase and had a cubic perovskite
structure at 300~K. The decrease in the lattice parameter in these samples compared with undoped
SrTiO$_3$ indicates the formation of a solid solution~\cite{PhysSolidState.61.390}.

\begin{figure}
\centering
\includegraphics{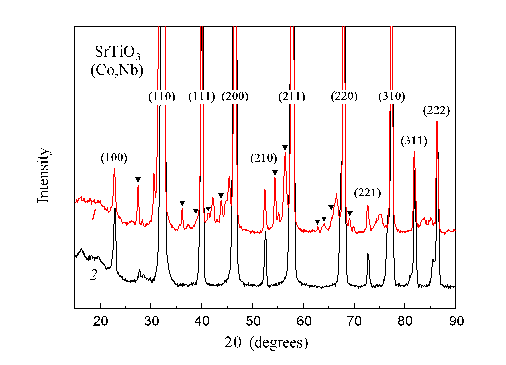}
\caption{Diffraction patterns of the SrTi$_{0.91}$Co$_{0.03}$Nb$_{0.06}$O$_3$ sample annealed
at (1) 1100$^\circ$C and (2) 1500$^\circ$C. Triangles indicate peaks of the TiO$_2$ phase.}
\label{fig1}
\end{figure}

X-ray studies of SrTiO$_3$ samples simultaneously doped with Co and Nb revealed that they also had
a cubic perovskite structure at 300~K (Fig.~\ref{fig1}). SrTi$_{0.91}$Co$_{0.03}$Nb$_{0.06}$O$_3$
samples annealed at 1500$^\circ$C were single-phase. In samples of the same composition annealed at
1100$^\circ$C, additional peaks were observed in the diffraction patterns; their identification using
the PCPDFWIN database indicates that most of them belong to the rutile TiO$_2$ phase (triangles next
to the curves). A few peaks, specifically the peak near $2\theta = 31^\circ$, could not be identified.
The lattice parameter of the SrTi$_{0.91}$Co$_{0.03}$Nb$_{0.06}$O$_3$ sample annealed at 1500$^\circ$C
is 3.9032$\pm$0.0003~{\AA}, which is 0.01~{\AA} larger than that of the sample doped with cobalt
only~\cite{PhysSolidState.61.390}.

\subsection{XANES spectra}

\begin{figure*}
\centering
\includegraphics{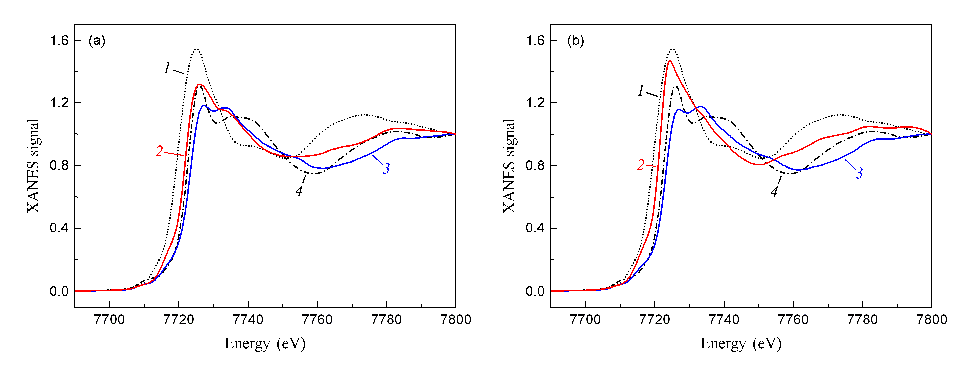}
\caption{XANES spectra of (2) SrTi$_{0.91}$Co$_{0.03}$Nb$_{0.06}$O$_3$ and (3) SrTi$_{0.97}$Co$_{0.03}$O$_3$
samples annealed at  (a) 1100 and (b) 1500$^\circ$C. Curves 1 and 4 are spectra of CoNb$_2$O$_6$ and
LaCoO$_3$ reference compounds.}
\label{fig2}
\end{figure*}

To determine the oxidation state of the Co impurity in SrTiO$_3$ samples, the positions of the 
absorption edge in the XANES spectra of strontium-enriched, nonstoichiometric SrTiO$_3$(Co) samples
and samples simultaneously doped with Co and Nb were compared with the edge positions in the
reference compounds CoNb$_2$O$_6$ (divalent cobalt) and LaCoO$_3$ (trivalent cobalt). The XANES
spectra of the samples annealed at 1100$^\circ$C and 1500$^\circ$C and reference compounds are
shown in Fig.~\ref{fig2}.
 
It can be seen that the absorption edges in the SrTi$_{0.97}$Co$_{0.03}$O$_3$ samples do not depend on
the annealing temperature and nearly coincide with the absorption edge of the LaCoO$_3$ reference 
compound. This suggests that the Co impurity in these samples has an oxidation state close to +3.

In SrTi$_{0.91}$Co$_{0.03}$Nb$_{0.06}$O$_3$ samples, the absorption edge shifts toward lower energies,
and the shift is more pronounced after annealing at 1500$^\circ$C. The difference between the XANES
curves for the samples annealed at 1100$^\circ$C and 1500$^\circ$C confirms that the samples annealed
at 1100$^\circ$C had not reached thermodynamic equilibrium. The shift in the curves indicates the
reduction of cobalt upon niobium doping, resulting in an oxidation state close to +2.

\subsection{Results of first-principles calculations}
\label{Sec3.3}

To construct structural models for the EXAFS data analysis, we modeled the structure of various
potential cobalt-containing centers with differing cobalt valences using first-principles
calculations. During the EXAFS data analysis, this approach makes it possible to rule out atomic
configurations that yield unphysical results or contradict other (magnetic, optical, or electrical)
properties of the samples under study. Thus, the combined analysis of EXAFS spectra and
first-principles calculations enhances the reliability of impurity-center identification.

The modeling of the impurity centers containing cobalt atoms located at the $A$~site (the Sr site) of
the perovskite structure was previously carried out in Refs.~\cite{PhysSolidState.61.390,JAlloysComp.820.153243}.

\begin{table*}
\caption{\label{table1}Local environment of Co ions in different theoretical models according to first-principles 
calculations. Average distance values are given for distorted structures; exceptions are distortions in 
the first shell of Jahn--Teller centers.}
\begin{ruledtabular}
\begin{tabular}{lcccccc}
Model & Magnetic \     & \multicolumn{5}{c}{Shell} \\
\cline{3-7}
      & moment~$S$     & 1 & 2 & 3 & 4 & 5 \\
\hline
Isolated Co$^{3+}$ ion at the $B$ site & 0 & 1.905 & 3.352 & 3.801 & 4.355 & 5.509 \\
(compensated by F)                     &   & (6O)  & (8Sr) & (6Ti) & (24O) & (12Ti) \\
Co$^{3+}$--nearest~V$_{\rm O}$--Rb$_{\rm Sr}$ & 1 & 1.909 & 3.398 & 3.897 & 4.351 & 5.511 \\
Co$^{3+}$--Nb$_{\rm Ti}$                      & 0 & 1.932 & 3.254 & 3.902 & 4.364 & 5.518 \\
Co$^{3+}$--nearest~V$_{\rm O}$--Co$^{3+}$     & 1 & 1.877 & 3.347 & 3.871 & 4.338 & 5.474 \\
Co$^{2+}$--distant~V$_{\rm O}$ (LS,           & 1/2 & 4$\times$1.929+ & 3.325 & 3.810 & 4.380 & 5.515 \\
Jahn--Teller distortion)                      &     & 2$\times$2.025  & \\
Co$^{2+}$--nearest~V$_{\rm O}$                & 1/2 & 1.929 & 3.377 & 3.862 & 4.338 & 5.512 \\
Co$^{2+}$--2Nb$_{\rm Ti}$ (HS)                & 3/2 & 2.030 & 3.315 & 3.883 & 4.394 & 5.541 \\
Co$^{2+}$--2Nb$_{\rm Ti}$ (LS,                & 1/2 & 4$\times$1.944+ & 3.301 & 3.859 & 4.383 & 5.533 \\
Jahn--Teller distortion)                      &     & 2$\times$2.078  & \\
\end{tabular}
\end{ruledtabular}
\end{table*}

Modeling of impurity centers with trivalent cobalt at the $B$~site (the Ti site) implies the presence
of defects that compensate for the charge difference between Co$^{3+}$ and Ti$^{4+}$ ions.
Calculations for an isolated Co$^{3+}$ ion at the $B$~site, where an extra electron was supplied
by a distant donor F atom at the oxygen site (see Table~\ref{table1}), indicated that the low-spin
state ($S = 0$) has the lowest energy. However, this result contradicted the experimentally observed
paramagnetism of cobalt-doped samples; therefore, we considered the Co$^{3+}$ complex with a
nearest-neighbor oxygen vacancy to be a more appropriate model for the impurity center~\cite{PhysSolidState.61.390}.
A comparison of the calculated distance of 1.909~{\AA} for this model (Table~\ref{table1}) with the
distance of $1.919 \pm 0.015$~{\AA} obtained from the EXAFS data analysis (Table~\ref{table2}) shows
good agreement between the two values.

Modeling of divalent cobalt impurity centers at the $B$~site, formed through charge compensation
by niobium impurities or oxygen vacancies, showed that different centers are characterized by different
Co--O distances in the first shell. Furthermore, these centers exhibit different magnetic moments.
Notably, the high-spin state of an isolated Co$^{2+}$ ion arises only when it is compensated by two
Nb donor atoms; when the Co$^{2+}$ ion is compensated by a distant oxygen vacancy, the high-spin
configuration relaxes into a low-spin one. It should be noted that for an isolated Co$^{2+}$ ion
compensated by two Nb$_{\rm Ti}$ atoms, two stable configurations are possible: a high-spin one and
a low-spin one. The latter configuration is characterized by a weak Jahn--Teller distortion accompanied
by elongation of the octahedron. Of these two configurations, the high-spin state is energetically more
favorable. For Co$^{2+}$ complexes with oxygen vacancies, the configuration with a nearest oxygen
vacancy is energetically far more favorable (by 529~meV) than the one with a distant vacancy. An
isolated Co$^{2+}$ ion compensated by a distant oxygen vacancy also exhibits a weak Jahn--Teller
distortion. Co$^{3+}$ ions, which appear at low Nb donor concentrations, possess a zero magnetic
moment.

Due to the slight systematic underestimation of interatomic distances characteristic of the LDA
approximation, the distances listed in Table~\ref{table1} were corrected for comparison with experimental
data by applying a scaling factor equal to the ratio of the experimental lattice parameter of bulk
SrTiO$_3$ to the calculated one. Since the cubic symmetry of supercells is almost inevitably reduced
upon the introduction of impurities, the table lists the average distance to the surrounding atoms
within the given shell in such cases.

\subsection{EXAFS spectra}

To determine the structural position and local environment of the Co impurity in SrTiO$_3$,
an analysis of the EXAFS spectra was performed. The oxidation state determined from XANES
measurements was taken into account during the data analysis and in selecting the structural
model.

\begin{figure*}
\centering
\includegraphics{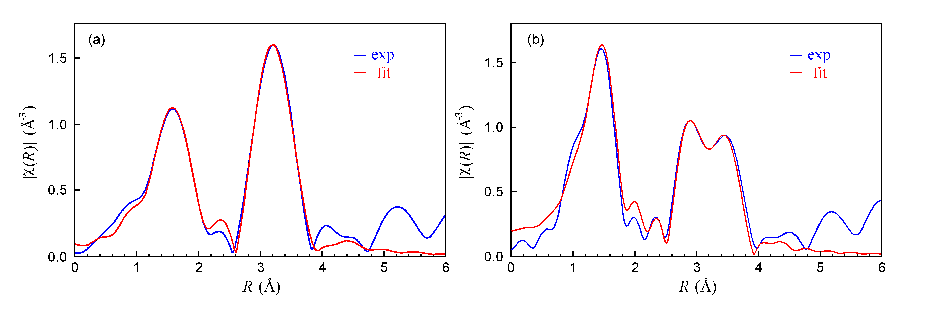}
\caption{\label{fig3}Comparison of Fourier transforms of the experimental (exp) and fitted (fit)
EXAFS spectra for (a) SrTi$_{0.91}$Co$_{0.03}$Nb$_{0.06}$O$_3$ sample annealed at 1500$^\circ$C and
(b) SrTi$_{0.97}$Co$_{0.03}$O$_3$ sample annealed at 1100$^\circ$C. Spectral fitting was performed
in the 0.8--4.1~{\AA} range.}
\end{figure*}

Fig.~\ref{fig3} shows a comparison of the Fourier transform amplitudes of the experimental and
calculated EXAFS spectra for the SrTi$_{0.91}$Co$_{0.03}$Nb$_{0.06}$O$_3$ sample annealed
at 1500$^\circ$C and the SrTi$_{0.97}$Co$_{0.03}$O$_3$ sample annealed at 1100$^\circ$C. We
selected the SrTi$_{0.97}$Co$_{0.03}$O$_3$ sample annealed at 1100$^\circ$C to minimize
possible cobalt incorporation into the $A$ sites. Unfortunately, the
SrTi$_{0.91}$Co$_{0.03}$Nb$_{0.06}$O$_3$ sample annealed at 1100$^\circ$C was not single-phase;
therefore, we used the sample annealed at 1500$^\circ$C for comparison. XANES data indicate that
Co atoms in these samples are in different oxidation states. A comparison of the EXAFS spectra
of these samples reveals qualitative differences between them.

For all SrTi$_{0.97}$Co$_{0.03}$O$_3$ samples containing trivalent cobalt~\cite{JAlloysComp.820.153243},
good agreement between the experimental spectra and the calculated curves was obtained for both the model
in which the Co$^{3+}$ ion substitutes for Ti$^{4+}$ with the formation of a distant oxygen vacancy and
the model in which the Co$^{3+}$ ion substitutes for Ti$^{4+}$ with the formation of a nearest-neighbor
oxygen vacancy. Our preference for the second model (Fig.~\ref{fig3}(b)) is based on the
calculations presented in Sec.~\ref{Sec3.3}, which indicate the absence of a magnetic moment in the
first model, contradicting the experimentally observed paramagnetism. Furthermore, although the
distances to the shells differ only slightly, the $S^2_0$ parameter, which accounts for many-electron
effects and inelastic scattering, is closer to the values found for reference compounds in the second
model.

\begin{table*}
\caption{\label{table2}Parameters of the local environment of cobalt atoms obtained from the EXAFS
data analysis. $R_i$ is the distance to the $i$th shell, $\sigma_i$ is Debye--Waller factor for
this shell, $S_0^2$ is the factor accounting for many-body and inelastic scattering effects at the
central atom.}
\begin{ruledtabular}
\begin{tabular}{cccc}
Parameter & First shell & Second shell & Third shell \\
\hline
\multicolumn{4}{c}{Sample with 3\% Co and 6\% Nb annealed at 1500$^\circ$C} \\
\hline
$S_0^2$   &  \multicolumn{3}{c}{0.791$\pm$0.314} \\
$R_i$     & 2.026$\pm$0.028 & 3.386$\pm$0.034 & 3.876$\pm$0.052 \\
$\sigma^2_i$ & 0.0065$\pm$0.0052 & 0.0079$\pm$0.0050 & 0.0088$\pm$0.0054 \\
\hline
\multicolumn{4}{c}{Sample with 3\% Co annealed at 1100$^\circ$C} \\
\hline
$S_0^2$   & \multicolumn{3}{c}{0.898$\pm$0.136} \\
$R_i$     & 1.919$\pm$0.015 & 3.371$\pm$0.014 & 3.886$\pm$0.014 \\
$\sigma^2_i$  & 0.0055$\pm$0.0025 & 0.0055$\pm$0.0016 & 0.0026$\pm$0.0014 \\
\end{tabular}
\end{ruledtabular}
\end{table*}

An analysis of the EXAFS spectra reveals that in SrTi$_{0.91}$Co$_{0.03}$Nb$_{0.06}$O$_3$ samples,
Co$^{2+}$ ions occupy Ti sites and are on-center (i.e., they occupy lattice sites without displacement).
The Co--O distance in the first shell increases compared with Nb-free samples ($R_{\rm Co-O} = 2.025$~{\AA},
see Table~\ref{table2}), a finding that is consistent with the conclusion about the change in the impurity
oxidation state drawn from the XANES spectra. As for the distances to the second and third shells,
they decrease (Table~\ref{table2}). A comparison of the obtained distances with those predicted by the
models discussed in Sec.~\ref{Sec3.3} shows that the only model that describes the experimental data is
that of an isolated Co$^{2+}$ ion with charge compensation provided by two Nb atoms. The model of a
Co$^{2+}$ ion with a nearest-neighbor vacancy is characterized by a shorter Co--O distance.

\section{Discussion}

In earlier papers discussing the properties of cobalt-doped SrTiO$_3$, it was assumed that oxygen
vacancies, which are formed to compensate for the difference in charge between Ti and Co ions,
play a significant role in determining these properties. Our study reveals a qualitative change
in the local environment of cobalt upon niobium doping: specifically, the disappearance of
nearest-neighbor oxygen vacancies. Calculations of the energies of different complexes
(Sec.~\ref{Sec3.3}) indicate that it is energetically more favorable for vacancies to be located
adjacent to cobalt atoms. Under our annealing conditions, thermodynamic equilibrium between free and
bound vacancies is established within minutes~\cite{JPhysChemC.126.4415}, a period much shorter than
the actual annealing times. Thus, the absence of vacancies in the local environment of Co implies
a very low concentration of free vacancies in the bulk of the sample. Consequently, with
regard to the doping strategy, co-doping with niobium proves to be a more effective method of
compensating the oxidation state of cobalt than compensation by vacancies. Apparently, the formation
energy of an oxygen vacancy exceeds the energy required to incorporate two Nb atoms. The practical
implication of this result is that niobium doping should reduce ionic conductivity, thereby negatively
affecting the fuel cell performance. This means that the question of optimal doping in such materials,
which typically incorporate both cobalt and donor impurities, requires careful consideration.

Another interesting effect observed in the results is a systematic decrease in the distances to the
second and third shells, in contrast to the increase in the first-shell distance. This effect appears
in both the calculations and the experiment, within the limits of measurement error. This observation
is in apparent contradiction with the increase in the experimental and calculated lattice parameter
(for the model with two Nb atoms, the lattice parameter increases by 0.012~{\AA} compared with
undoped SrTiO$_3$). The calculated increase in the distance to the first shell is approximately
0.07~{\AA}, while the decreases for the second and third shells are 0.06~{\AA} and 0.03~{\AA},
respectively.

\begin{figure}
\centering
\includegraphics{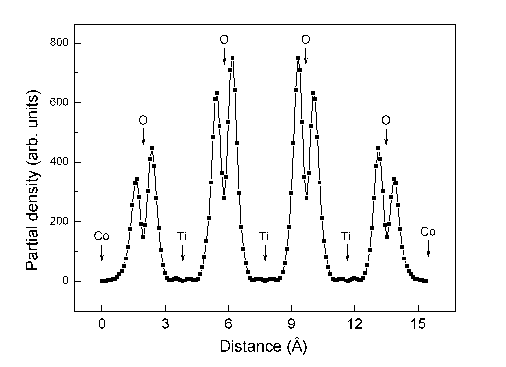}
\caption{\label{fig4}Dependence of the partial electron density of the valence states for the
Co--O--Ti--O--Ti--O--Ti--O--Co chain oriented along the [001] axis. Arrows indicate the positions
of the atoms in the chain.}
\end{figure}

It is known that the chemical bonding in SrTiO$_3$ has mixed ionic and covalent character. To clarify
the roles of these two bonding types in the observed phenomenon, we calculated the electron-density
distribution along the fourfold axes extending from the cobalt ion in a crystal co-doped with Co and
Nb. Fig.~\ref{fig4} shows the partial electron density obtained by averaging the squared moduli of
the wave functions of twenty branches of the electronic spectrum near the top of the valence band at
the $\Gamma$~point over the three directions of the cubic lattice. These branches make the dominant
contribution to chemical bonding. The electron-density maxima are seen to be concentrated in the
$p$~orbitals of the oxygen atoms. An important feature of this distribution is the asymmetry of the
electron density on the $p$ orbitals. We attribute this asymmetry to the polarizing effect of Nb atoms
on the electric-field distribution in the unit cell. Indeed, the regions near the Co$^{2+}$ ions in our
structure carry a net negative charge relative to that near the regular Ti$^{4+}$ ions, whereas the
regions near the Nb$^{5+}$ ions, which are located at the maximum distance from the Co ions, carry
a positive charge. The electrons in the oxygen $p$~orbitals respond to the resulting electric field
and partially screen it through redistribution within the $p$~orbitals. Fig.~\ref{fig4} shows that the
covalent character responsible for the anomalous increase in the Born effective charges in SrTiO$_3$
manifests itself not as the appearance of electron-density peaks near the centers of chemical bonds,
as observed in silicon, but rather as a redistribution of charge density on the oxygen $p$~orbitals.
This implies that the actual picture of ferroelectric instability in perovskite crystals is closer to
the model of nonlinearly highly polarizable oxygen ions proposed previously in~\cite{PhysRevB.35.4840}.

The electron-density distribution shown in Fig.~\ref{fig4} indicates that the change in electron density 
along the Ti--O bond for the oxygen atom closest to the cobalt atom in the chain is relatively small.
This suggests that the covalent-bond contribution to the observed shortening of the corresponding
Ti--O bond length (by 0.1~{\AA}) is not significant. On the other hand, the electrostatic attraction
between the positively charged Sr$^{2+}$ and Ti$^{4+}$ ions and the region of local negative charge
around the Co atom may be strong enough to cause the observed reduction in the corresponding
interatomic distances. Thus, we conclude that the Coulomb interaction between the ions and the
electrostatic-potential modulations induced by the incorporation of non-isoelectronic impurities into
the crystal is the primary cause of the static atomic displacements observed in both calculations and
experiment.

Calculations previously performed by other authors have shown that divalent cobalt is required for
the emergence of room-temperature ferromagnetism in SrTiO$_3$(Co). Our calculations indicate that the
magnetic moment of such an ion can be even higher than that of a trivalent ion in a complex with an
oxygen vacancy (see Table~\ref{table1}). Surprisingly, the highest magnetic moment, $S = 3/2$, is
found for divalent cobalt ions charge-compensated by Nb ions, whereas a divalent cobalt complex with
a nearest vacancy has a magnetic moment of $S = 1/2$. The mechanisms underlying the stabilization of
low-spin and high-spin states for trivalent cobalt ions were previously discussed in~\cite{JApplPhys.36.879},
where it was shown that these states arise from competition between Hund's exchange energy and
crystal-field strength. Since crystal-field strength depends on the distance between the Co atom and
the nearest ligands, changes in the local environment can alter the atom's magnetic moment. Indeed,
theoretical calculations of the effect of strain on the magnetic state of the Co$^{3+}$ ion in
LaCoO$_3$~\cite{SciRep.4.7003} confirm this effect. According to our calculations, which agree with
the experimental data, the lattice parameter of SrTiO$_3$ decreases upon the introduction of oxygen
vacancies and increases upon niobium doping. Consequently, an increase in the Co--O distance weakens
the crystal field, causing the $d_{x^2-y^2}$$\uparrow$ level to drop below the $d_{xz}$$\downarrow$
and $d_{yz}$$\downarrow$ levels, thereby making the high-spin state energetically more favorable.
This explains the nature of the observed change in the magnetic moment of the Co$^{2+}$ ion.

\section{Conclusions}

XAFS spectroscopy and first-principles calculations were used to perform a detailed study of
SrTiO$_3$ samples co-doped with cobalt and niobium. These studies provided reliable data
on the local structure and oxidation state of the Co impurity at the $B$~site of SrTiO$_3$,
depending on the synthesis conditions and niobium doping. Simultaneous doping with Co and Nb
leads to a reduction in the cobalt oxidation state to +2, accompanied by an increase
in the distance between the impurity atom and the nearest oxygen atoms, consistent with
first-principles calculations. Furthermore, it was revealed that niobium doping inhibits the
formation of oxygen vacancies. First-principles calculations demonstrated that the Coulomb
interaction between ions and local electric fields induced by non-isovalent impurities are
the primary drivers of the local atomic distortions observed experimentally in the unit cell.
Besides, the competition between the exchange interaction and crystal-field effects can
potentially lead to a significant enhancement of the Co magnetic moment as the lattice
parameter increases.

\begin{acknowledgments}
The authors express their gratitude to the BESSY staff for the opportunity to perform the 
measurements.
\end{acknowledgments}


\providecommand{\BIBYu}{Yu}

\end{document}